\documentclass[12pt]{iopart}
\usepackage{iopams}
\usepackage{epsf}
\usepackage[dvips]{graphicx}
\begin{document}
\title{A class of generalised Jordan-Schwinger maps}
\author{ N. M. Oliveira-Neto$^{1,2}$, E. M. F. Curado$^{1}$ and M. A. Rego-Monteiro$^{1}$}
\address{$^{1}$ Centro Brasileiro de Pesquisas F\'\i sicas,\\
Rua Xavier Sigaud,150, 22290-180,\\Rio de Janeiro, Brazil.\\
$^{2}$ Departamento de F\'\i sica \\
Universidade Federal de Vi\c{c}osa, \\ 36570-000, Vi\c{c}osa, Brazil.}
\ead{\mailto{nmon@ufv.br},\mailto{evaldo@cbpf.br},\mailto{regomont@cbpf.br}}

\begin{abstract}
In this article we introduce a class of generalisations of the Jordan-Schwinger (JS) 
map which realises the recent proposed generalised $sl(2)$ (G-$sl(2)$) algebra via two 
independent Generalised Heisenberg Algebras (GHA). Although the GHA and
the G-$sl(2)$ algebra exhibit more general 
algebraic structures than the Heisenberg and $sl(2)$ algebras, 
the generalised JS map presents a compact and simple structure wich
includes the standard JS map as a particular case. Finally, since in the GHA 
there is a physical interpretation in terms of composite particles, we will carry out this 
assertion in a manner that the generalised $sl(2)$ algebra could be related to
composite particles with angular momentum. 
\end{abstract}

\section{Introduction}
A realisation of a given set of operators via Bose creation and annihilation
operators is an importante tool for solving a large number of quantum problems \cite{palev}. 
The Jordan-Schwinger (JS) map \cite{jordan,schwinger}
is a very interesting and widely known bosonisation of the Lie algebra which
connects the angular momentum algebra 
to the two independent Heisenberg algebras \cite{man'ko}. It
has proven itself a useful tool in physics \cite{mota,klein1},
as well as its generalised version \cite{man'ko,daskaloyannis,palev}.
This method has been extended to the classical (super)
Lie algebras by employing various types of oscillator algebras such as
Fermi, para-Bose, and para-Fermi \cite{daskaloyannis,kyung}.
As in the classical Lie algebra theory, the JS method has been also
used \cite{man'ko,biedenharn,macfarlane} to realise
the $su_{q}(2)$ algebra whose generators are written as a quadratic
form of the generators of the $q$-oscillator algebra
\cite{biedenharn,macfarlane}. The $su_{q}(2)$ and the $q$-oscillator algebras are deformed
versions of the $su(2)$ and Heisenberg algebras, respectively.

Due to the fact that the Heisenberg and $su(2)$ algebras are paradigmatic structures 
\cite{curado}, there were been efforts in order to generalise these algebraic 
structures and to seek possible applications in diverse areas of 
physics \cite{curado,faddev}.

In this sense, two classes of algebraic structures called Generalised Heisenberg Algebra
(GHA) \cite{curado2} and Generalised $sl(2)$ (G-$sl(2)$) algebra \cite{curado,curado1} were
introduced generalising the Heisenberg and $sl(2)$ algebras, respectively. 
These structures are constructed with, in a general case, different characteristic functions of one of
their generators. When these functions are linear with slope $q=1$, they recover
the standard Heisenberg and $sl(2)$ algebras
\cite{curado2,curado1}; when the functions are linear with
slope $q \ne 1$, the algebras turn into the $q$-oscillator and $sl_{q}(2)$
algebras. In the case of non-linear characteristic functions, the 
GHA and the G-$sl(2)$
algebra present more general algebraic structures than the $q$-oscillator and $sl_{q}(2)$ 
algebras, respectively.

Concerning its physical content, the GHA has two different interpretations. One
interpretation allows one to consider the formalism of second quantisation
for composite particles \cite{curado3,curado4,ribeiro-silva1}. By the other
hand, the GHA can be used in order to describe phenomenologicaly the energy spectrum of composite particles 
\cite{desouza1,oliveira-neto}. 

The aim herein is to realise the JS method in order to connect 
the GHA to the G-$sl(2)$ algebra in the case of non-linear
characteristic functions present in each algebra. Motivated by the
interpretation of the GHA in terms of composite 
particles, we will carry out this assertion, 
as it will become clear later, in such a way that the 
G-$sl(2)$ algebra could be related to composite particles with angular
momentum. 
 It is worth to stress that this class of generalised JS maps that connect 
the GHA to the G-$sl(2)$ algebra for non-linear characteristic functions
will be introduced with a simple and compact structure which contains
the standard JS map as a particular case.

In section 2 we will discuss the GHA for a specific non-linear characteristic 
function. In section 3, we will discuss the G-$sl(2)$ algebra for another
corresponding 
specific non-linear characteristic function. In section 4 we will present the 
generalised JS map which connects the GHA to the G-$sl(2)$ algebra.
Section 5 are the conclusions.

\section{Generalised Heisenberg Algebra}

Recently, it was constructed an algebraic structure that is a 
generalisation of the Heisenberg algebra \cite{curado2}.
This structure is generated by the Hamiltonian $H$ and the ladder operators
$A$ and $A^{\dagger}$ satisfying the following relations
\begin{eqnarray}
H \,  A^\dagger &=& A^\dagger \,  f(H) \label{eq:1}\\
A  \, H &=& f(H) \,  A \label{eq:2}\\
\left[ A , A^\dagger \right] &=& f(H)-H  \label{eq:3}
\end{eqnarray}
where, by hypothesis, $(A^{\dagger})^{\dagger}=A$, $ H^\dagger=H
$ and $f(H)$ is an analytic function of $H$. $f(H)$ is known as
characteristic function of the algebra. Using the relations 
(\ref{eq:1}-\ref{eq:3}) we can see \cite{curado2} that the operator
\begin{eqnarray}
C \equiv A^\dagger  \,  A - H = A  \, A^\dagger  - f(H) \label{eq:4}
\end{eqnarray}
is the casimir operator of the algebra.
Starting from the vacuum state $|0\rangle$ of $H$, $H|0\rangle \equiv
\alpha_0 |0\rangle$, we can define the
basis of the Fock space by applying successively the operator
$A^{\dagger}$ on the vaccum state $|0\rangle$. The representations
of the algebra are obtained by the action of the GHA generators on this basis as  
\begin{eqnarray}
H \, |m\rangle &=& f^{(m)}(\alpha_0) \, |m\rangle,\,\,\, 
m=0,1,2,... \label{eq:5} \\
A^\dagger \, |{m}\rangle &=& M_{m} \, |m+1\rangle \label{eq:6} \\
A \, |{m}\rangle &=& M_{m-1} \, |{m-1}\rangle, \label{eq:7}
\end{eqnarray}
where $M^2_{m-1} = f^{(m)}(\alpha_0) - \alpha_0$
and $f$ is the characteristic function whose 
$m$-th iteration of $\alpha_0$ through function $f$ is denoted by 
$f^{(m)}(\alpha_0)$.  

Considering a linear characteristic function $f(x)=x+1$ in relations
(\ref{eq:1}-\ref{eq:3}), we recover the standard commutation
relations of the Heisenberg algebra \cite{curado2}. If it is a linear function with
slope  $q \ne 1$, i.e. $f(x)=q x+1$, the GHA corresponds to the
$q$-oscillators algebra \cite{curado2}; finally, considering a
non-linear characteristic function $f(x)$, there is not a direct
relation between the GHA and the $q$-oscillator algebra, showing that this
formalism allows us to obtain more general algebraic structures than
the $q$-oscillator and Heisenberg algebras.
For this reason the GHA is a class of generalised Heisenberg
algebras since from it we can recover the deformed ($q$-oscillator algebra) and
non-deformed (Heisenberg algebra) Heisenberg algebras \cite{curado1}. 

It is worth to stress that this class of generalisation of Heisenberg-type algebras 
describes a family of one-dimensional quantum systems having any two
successive energy levels related by the relation $\varepsilon_n =
f(\varepsilon_{n-1})$, where $f$ is the characteristic function of 
the algebra, which is different for each physical system \cite{curado5}
and non-linear in general. It was also shown that it is possible to use the GHA as a linear
characteristic function to construct a
non-standard Quantum Field Theory \cite{curado3,curado4}. In fact, this is possible even
if the characteristic function
is a non-linear one \cite{ribeiro-silva1}, in which case we can overcome some
limitations in order to fit experimental data. An other case in which the
GHA with a non-linear characteristic
function works better than the GHA with a linear characteristic
function is in molecular
physics \cite{desouza1}. In this case, it was necessary to change from the 
$q$-oscillator algebra (or, equivalently, the GHA with a linear characteristic function) 
in order to reproduce faithfully the
experimental data of the energy levels of the carbon monoxide molecule
(CO). So, the use of non-linear characteristic function seems to be a useful
tool in many different physical systems. 
It is about this sort of characteristic functions
that we are going to discuss in the following.   

For a given characteristic function $f$, the representations
of the algebra can be studied by an analysis of the stability of the
attractors of $f$ and the attractors of their composed functions \cite{curado2}.
For shortness, since the central issue in this present paper is to connect the
G-$sl(2)$ algebra to the GHA with non-linear
characteristic functions, we will study the finite- and
infinite-representations of the GHA focusing on the graphical analysis
of the stability of the fixed points of a second degree polynomial equation 
$f(\alpha)=t\alpha^2+q\alpha+s$ with $t>0$. The fixed points of this function
are solutions of $f(\alpha^*)=\alpha^*$; herein we will work, for simplicity,
with $\Delta=(q-1)^2-4ts=0$ under the restriction  $\alpha > -q/(2t)$ ($f^{'}(-q/(2t))=0$). 
The restriction $\alpha > -q/(2t)$ is used in order to
have an inversible function.

In this case $f$ presents a fixed point
$\alpha^{*}=(1-q)/2t$. The choice $\alpha_0 = \alpha^{*}$ corresponds
to a one-dimensional representation of the algebra. 
There are also infinite-dimensional representations with
$\alpha_0$ belonging to the regions: $(a)$ $-q/(2t)< \alpha_0 < \alpha^* $ 
and $(b)$ $ \alpha^* < \alpha_0 < \infty$. In the region $(a)$ the $H$ 
eigenvalues go to the asymptotic value $\alpha^*$ while in $(b)$ they
go to infinity.

All these analyses can be more easily seen in figure 1, where we plot
the functions $f(\alpha)=2.5x^2-2.5x+1.225$ and $y=\alpha$ versus $\alpha$.
In this figure, we exhibit the iterations of $\alpha_0^{a}=0.56$ and $\alpha_0^{b}=0.85$.

\begin{figure}
\begin{center}
\includegraphics*[angle=0,scale=0.35]{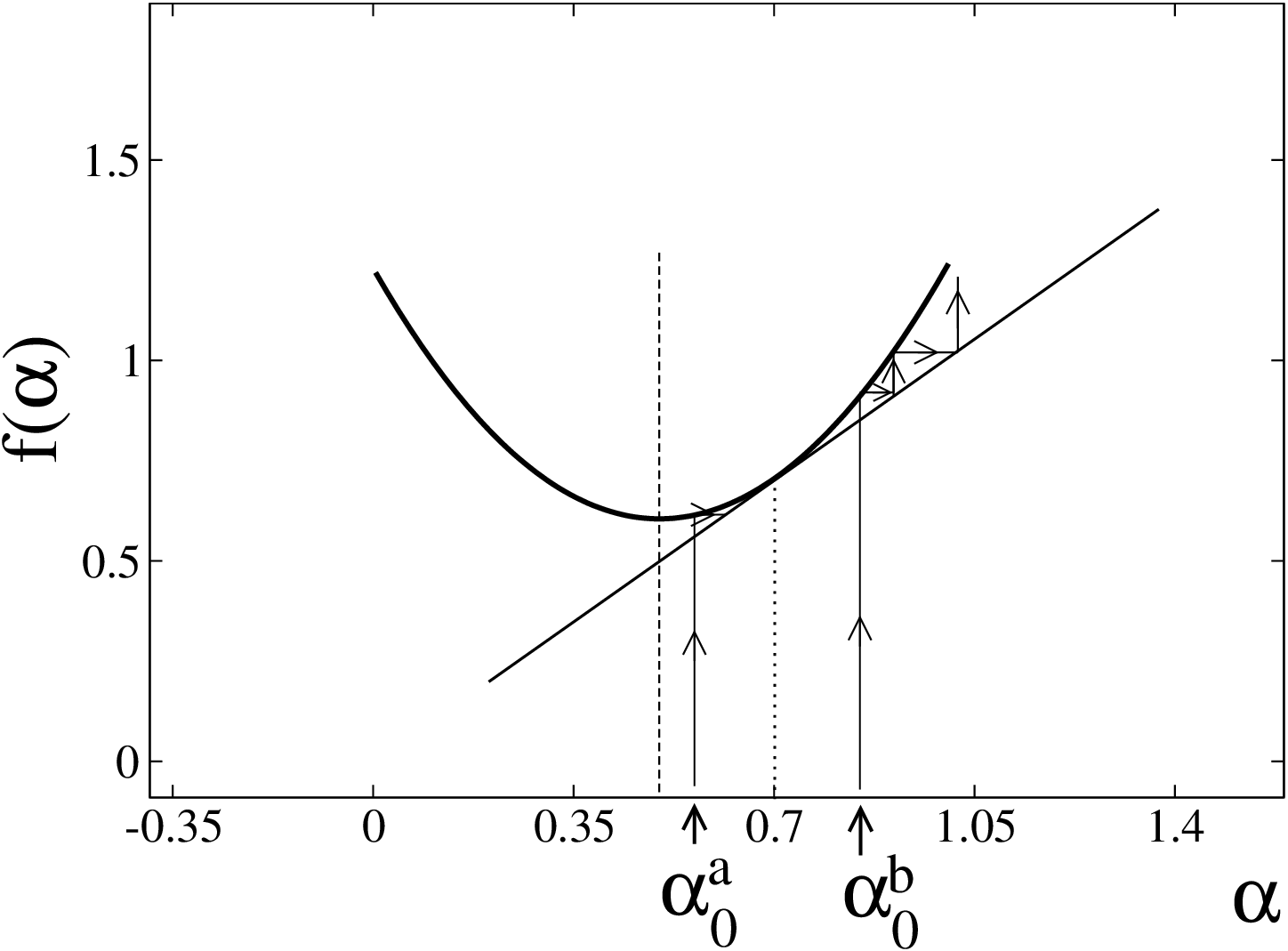}
\caption[Iterações de $\alpha_0$ através de uma $f$ quadrática
  ($\Delta=0$)] {\small{Graphical analysis of the representations of 
    the GHA for $f(\alpha)=2.5\alpha^2 -2.5 \alpha +1.225$ ($\Delta=0$). Note that
    there is one fixed point $\alpha^{*}=0.7$ (vertical doted line)
  within the region under analysis, $\alpha > -q/(2t)=0.5$ (vertical
  dashed line). Note also that $\alpha_0^{a}$ is a starting point
  whose future iterations tend to the fixed point $\alpha^{*}=0.7$ while
  $\alpha_0^{b}$ is a starting point whose future iterations go to
  infinity. These two different choices correspond to infinite-dimensional
  representations for the GHA while the choice $\alpha_0=\alpha^{*}$
  correspond to a one-dimensional representation of the algebra.
  }}
\label{Fig. 1}
\end{center}
\end{figure}

\section{Generalised $sl(2)$ algebra}

In a similar way that was constructed the GHA,
a generalisation of the $sl(2)$ algebra was introduced where in the
commutation relations of their generators there is a characteristic
function $g$ of one of their own generators 
\cite{curado1}. This new algebraic structure is constructed by the
following relations among their generators   
\begin{eqnarray}
J_0 \,  J_- &=& J_- \, g(J_0) \label{eq:13} \\
J_+ \, J_0 &=& g(J_0) J_+ \label{eq:14} \\
\left[ J_+,J_- \right] &=& J_0
(J_0 +1)-g(J_0)(g(J_0) +1) \label{eq:15} 
\end{eqnarray}
where $J_- =(J_+)^{\dagger}$, $J_0^{ \, \dagger}= J_0$ 
and $g(J_0)$ is any analytical function in 
$J_0$ \cite{curado1}. Using the
eqs.(\ref{eq:13}-\ref{eq:15}) it was shown that the operator
\begin{equation}
\widehat{C} \equiv \frac{1}{2} \left\lbrace {J_+ \, J_- + J_- }
  \, J_+  + J_0 (J_0 +1)+
g(J_0)(g(J_0) +1) \right\rbrace \label{eq:16}
\end{equation}
is the casimir operator of the algebra. If we chose the function
$g(J_0)= J_0 -1$ we can easily see that the relations 
(\ref{eq:13}-\ref{eq:15}) become the widely known commutation
relations of the $sl(2)$ algebra \cite{sakurai}.
Departuring from this special case for the characteristic function $g$ we 
obtain different algebraic structures that generalise the $sl(2)$
algebra in the sense that we can recover the standard $sl(2)$ algebra when
$g(J_0) \rightarrow  J_0 -1$. For instance, 
considering $g(J_0)= q J_0 -1$, the
G-$sl(2)$ algebra corresponds to the $sl_{q}(2)$ algebra, which in
the limit $q \rightarrow 1$ turns into the standard $sl(2)$
algebra \cite{curado1}. In the non-linear case, the algebraic
structure become more general than the $sl(2)$ and $sl_{q}(2)$
algebras. Is in this non-linear case for the characteristic function $g$ that we are going
to exhibit a class of generalised JS maps
that realises the generators of the G-$sl(2)$ algebra
in terms of the generators of the GHA with a non-linear
characteristic function $f$.

Seeking for the representations of the G-$sl(2)$ algebra, we
suppose \cite{curado1} that exists a highest weight state of the representation
$|\alpha_j;j\rangle$ defined as
\begin{equation}
J_+ \, |\alpha_j ; j\rangle \equiv 0 \label{eq:17}
\end{equation}
whose $J_0$ eigenvalue is $\alpha_j$, i.e.
$J_0 |\alpha_j ; j\rangle=\alpha_j |\alpha_j ; j\rangle $, where $\alpha_j$ is
a real number and $j$ is a natural number.
Applying sucessively $J_-$ on this highest weight state, 
we have different $J_0$ eigenstates that form a basis of the Fock space. The
representation of the algebra is obtained by the application of the
algebra generators on this basis. For a general characteristic
function $g$ we obtain, for $m=0,1,...,2j$, the following representation of the algebra 
\begin{eqnarray}
J_0 \, \, |\alpha_{j};j-m\rangle &=& \alpha_{j-m}
|\alpha_{j};j-m\rangle    \label{eq:18} \\
J_+ \, \, |\alpha_{j};j-m\rangle &=& \widehat{M}_{m-1}
|\alpha_{j};j-m+1\rangle    \label{eq:19} \\
J_- \, \, |\alpha_{j};j-m\rangle &=& \widehat{M}_{m}
|\alpha_{j};j-m-1\rangle    \label{eq:20} \\
\widehat{C} |\alpha_{j};j-m\rangle &=& \alpha_j (\alpha_j +1) \, 
|\alpha_{j};j-m\rangle \label{eq:21}
\end{eqnarray}
where by hypothesis the function $g$ and the initial value $\alpha_j$
satisfy $\alpha_j >g^{(m)}(\alpha_j)$, where $g^{(m)}(\alpha_j)$ is the
$m$-th iteration of $\alpha_j$
through $g$, $\widehat{M}^{2}_m = 
(\alpha_j - \alpha_{j-m-1})(\alpha_j +\alpha_{j-m-1}
+1)=\alpha_j(\alpha_j+1)-\alpha_{j-m-1}(\alpha_{j-m-1}+1)$ and 
$\alpha_{j-m}=g^{(m)}(\alpha_j)=g(\alpha_{j-m+1})$ for a integer $m \ge
1$ \cite{curado1}. 

Hereafter we are going to discuss the finite- and infinite-dimensional
representations of the algebra described by the commutation relations
(\ref{eq:13}-\ref{eq:15}), where we will consider
a non-linear charactereristc function $g(\alpha)=t^{'}\alpha^2
+q^{'}\alpha-s^{'}$ with $t^{'}<0$ and under the restriction
$\alpha<-q^{'}/(2t^{'})$ in order to consider inversible functions.

For the finite-dimensional representations, since each representation is
constructed from the highest weight state,
we can find out two different constraints for $\alpha_j$ such that
the eq.(\ref{eq:20}) is identically null, i.e. for a $d$-dimensional
representation $\widehat{M}_{d+1}=0$, given by
\begin{equation}
\alpha_j = \alpha_{j-d} \label{eq:23}
\end{equation}
and
\begin{equation}
\alpha_j + \alpha_{j-d} +1=0 . \label{eq:24}
\end{equation}
Otherwise, the dimension of the representations will be infinite.

The solutions of the eq.(\ref{eq:23}) can be studied by analysis of the
stability of the attractors of $g$ and their composed functions. 
So, we are going to focus on the graphical analysis of the stability of the fixed points
$\overline{\alpha^{*}}$, attractors of period 1, solutions of the equation
$\overline{\alpha^{*}}=g(\overline{\alpha^{*}})$. For simplicity, we will
consider the case $\Delta=(q^{'}-1)^2+4t^{'}s^{'}=0$.

\begin{figure}
\begin{center}
\includegraphics*[angle=0,scale=0.3]{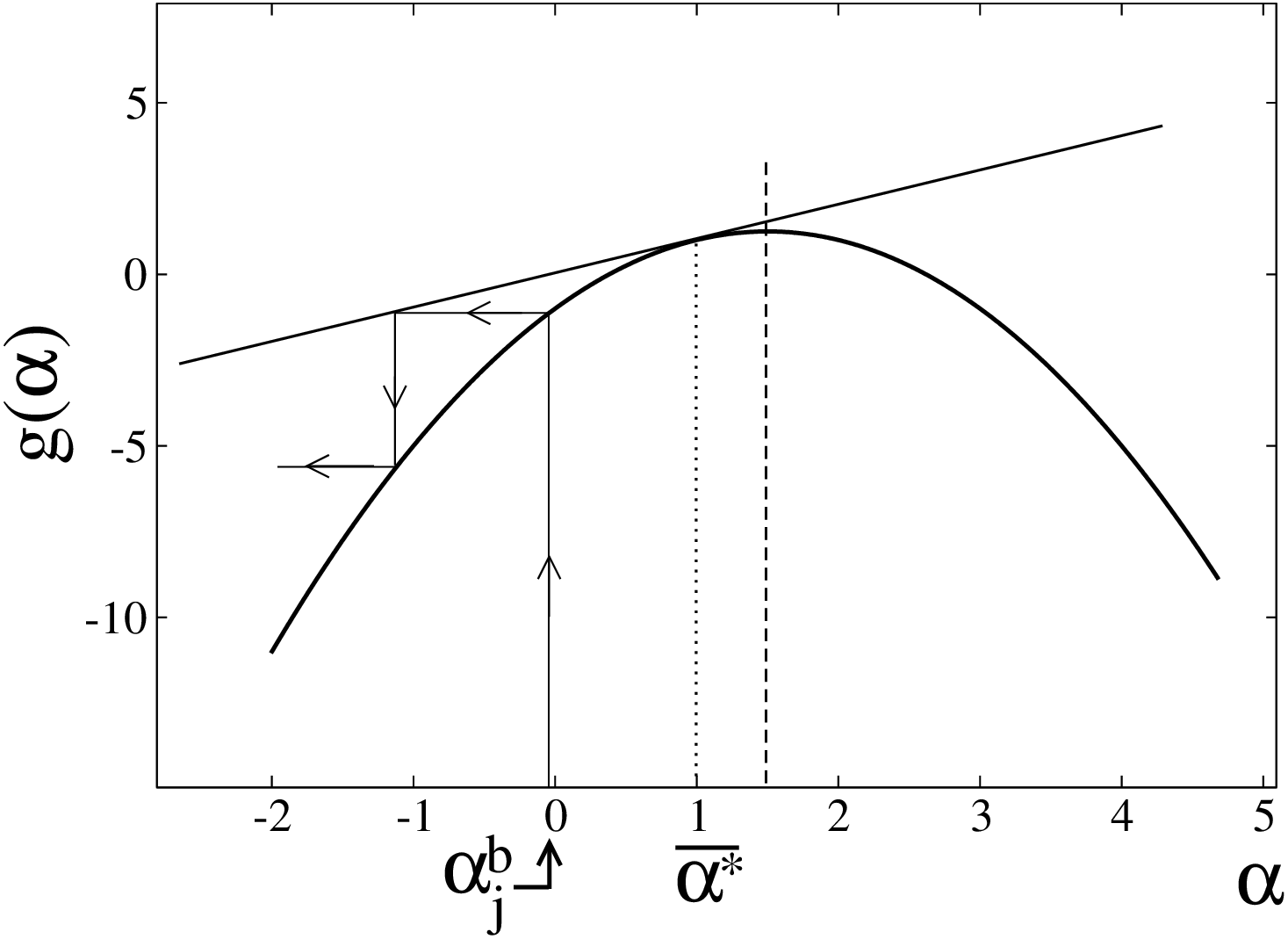}
\caption[]{\small{Graphical analysis of the representations of the 
generalised $sl(2)$ algebra for $g(\alpha)=- \alpha^2 +3 \alpha -1$
    ($\Delta=0$). Note that there is one fixed point $\overline{\alpha^{*}}=1$
 and there are two different regions that correspond to two different
 asymptotic behaviours to the $J_0$ eigenvalues associated with
 infinite-dimensional representations of the algebra. In the figure, we
 exhibit the iteration of $\alpha^{b}_j$ whose future
    iterations tend to $-\infty$. Note also that the one-dimensional
    representation can be obtained for the choice $\alpha_j=\overline{\alpha^{*}}$.

 }}
\label{Fig. 2}
\end{center}
\end{figure}

In this case $g$ presents a fixed point
$\overline{\alpha^{*}}=(1-q^{'})/2t^{'}$; for $\alpha_j
=\overline{\alpha^{*}}$ the algebra has a one-dimensional
representation ($d=1$) since
$\alpha_{j}=\alpha_{j-1}=\overline{\alpha^{*}}$.
Apart from this finite-dimensional representation, the algebra has 
infinite-dimensional representations associated with the following regions of
$\alpha_j$: ($a$) $\overline{\alpha^{*}} < \alpha_j < -q^{'}/(2t^{'})$
 and ($b$) $-\infty < \alpha_{j} < \overline{\alpha^*}$. 
In the case $(a)$ the $J_0$ eigenvalues, $g^{(m)}(\alpha_j)$, go to the
asymptotic value $\overline{\alpha^*}$ while in the region $(b)$ go to
$-\infty$.
In the figure 2 we plot the functions $g(\alpha)=- \alpha^2 +3 \alpha
-1$ and $y=\alpha$ versus $\alpha$ and exhibit the interations of the
$\alpha_j^{b}=-0.05$. The vertical doted line corresponds to the fixed point
$\overline{\alpha^*}= 1$ and the vertical dashed line corresponds to
$\alpha=-q^{'}/(2t^{'})=1.5$.

\begin{figure}
\begin{center}
\includegraphics*[angle=0,scale=0.3]{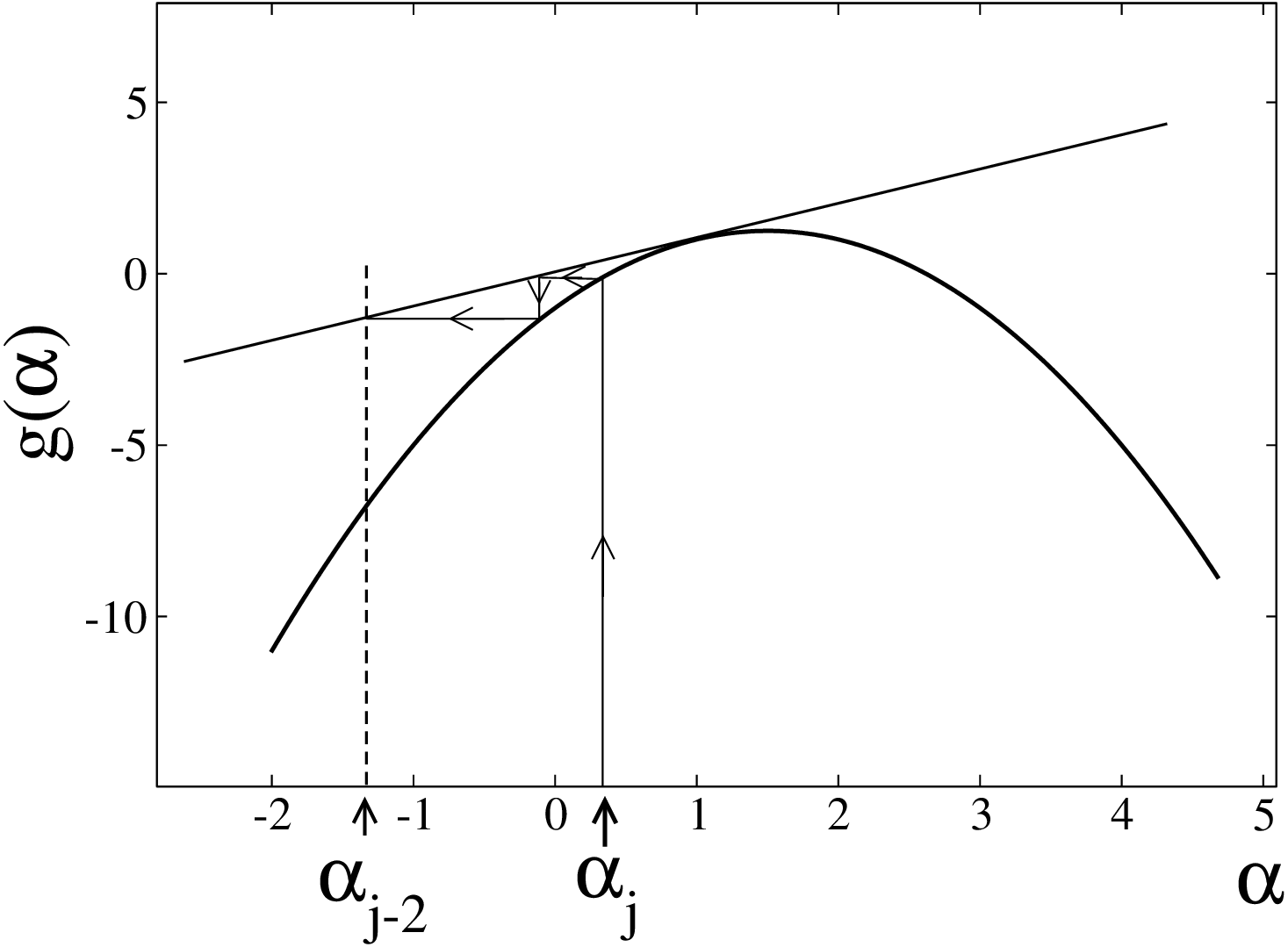}
\caption[]{\small{Graphical analysis of the representation of the generalised $sl(2)$
    algebra  for $g(\alpha)=- \alpha^2 +3 \alpha -1$
    ($ \Delta=0$). Note that there is a finite-dimensional
    representation associated with the cut condition equation,
    $\alpha_{j-2}=-\alpha_j -1$. In this plot we show a two-dimensional
    representation ($j=1/2$) with $\alpha_j=0.33479$. There is
    also another two-dimensional representation of the algebra for
    $\alpha_j=2.9228$, but this value is out of the region,
    $\alpha<-q^{'}/(2t^{'})= 1.5$, in analysis.
 }}
\label{Fig. 3}
\end{center}
\end{figure}

It is important to say that the algebra has 
infinite-dimensional representations if we choose an $\alpha_j$ within
 the allowed regions and if the cut condition equation (\ref{eq:24}) is not
satisfied for any future iteration. As stated before, the cut condition equation (\ref{eq:24})
corresponds to the $d$-dimensional representations of the algebra. 
For instance, using $t^{'}=-1$, $q=3$ and $s^{'}=1$, that corresponds
to $\Delta =0$, and selecting the representation
$j=1/2$ ($d=2$) we obtain $\alpha_j=0.33479$ and $\alpha_j=2.9228$
as solutions of the eq.(\ref{eq:24}). So, in order to obtain 
two representations of dimension $d=2$ of the algebra, we can choose
one or the other solution for $\alpha_j$.

An example of this finite dimensional case is showed in figure 3 where
we plot the function $g(\alpha)=-\alpha + 3\alpha -1$ and $y=\alpha$
versus $\alpha$ and where we select the two-dimensional representation $d=2$
($j=1/2$) choosing $\alpha_j=0.33479$. We also plot the vertical doted line
$\alpha_{j-2}=\alpha_j-1=-1.33479$, which represents the cut condition equation
(\ref{eq:24}). Since $\alpha_j=2.9228$ is out of the 
region in analysis, $\alpha<-q^{'}/(2t)=1.5$, we did not exhibit the
interations of this solution.

\section{A class of generalised Jordan-Schwinger maps}

First of all, we consider the algebra of two independent (uncoupled) 
deformed oscillators which we call the $1$-type and the $2$-type. We 
assume that the commutation
relations of the generators of their individual algebras 
 satisfy the relations (\ref{eq:1}-\ref{eq:3}) for 
a general analytical function $f$. As we saw in section 2, these algebras have a general 
representation given by
\begin{eqnarray}
H_{1} |n_{1},n_{2}\rangle &=& f^{(n_{1})}(\alpha_0) 
|n_{1},n_{2}\rangle \;
\:  \label{eq:27} \\ 
A^{\dagger}_1 |n_{1},n_{2}\rangle &=& M_{n_{1}} 
|n_{1}+1,n_{2}\rangle \label{eq:28} \\
A_{1} |n_{1},n_2\rangle  &=&  M_{n_1-1} |n_{1}-1,n_2\rangle 
\label{eq:29}
\end{eqnarray} 
where $M^{2}_{n_{1}-1}=f^{(n_1)}(\alpha_0)- \alpha_0$,
$n_{i}=1,2,3,....$ for $i=1,2$ and $|n_1,n_2\rangle \equiv |n_1\rangle
\otimes |n_2\rangle$, where $\otimes$  is the tensorial product.
Analogously $H_{2},A^{\dagger}_2$ and $A_2$ act on the second entry of $|n_1 , n_2\rangle$.
Here we consider that $H_{1}$ and $H_{2}$ have vacuum eigenstates with same
eigenvalues, i.e. $\alpha^{(1)}_0 =\alpha^{(2)}_0 \equiv \alpha_0$, where $H_{1}
|0,0\rangle = \alpha^{(1)}_0 |0,0\rangle$ and $H_{2} |0,0\rangle = \alpha^{(2)}_0 |0,0\rangle$.
  
Starting from the vacuum eigenstate $|0,0\rangle$ 
 we can construct different  
eigenstates of the two deformed oscillators given by
\begin{equation}
|n_1,n_2\rangle = \frac{(A^\dagger_1)^{n_1}(A^\dagger_2)^{n_2}}
{M^{n_1+n_2}_0\sqrt{\left[n_1\right]_{f}!
\left[n_2\right]_{f}!}} |0,0\rangle \label{eq:30}
\end{equation}
where the generalised Gauss number of $m$, denoted by $\left[
  m\right]_{f}$, for a general function $f$, is defined as \cite{curado2}
\begin{equation}
\left[m \right]_{f} \equiv \frac{M^2_{m-1}}{M^{2}_0}= \frac{f^{(m)}
(\alpha_0)-\alpha_0}{f(\alpha_0)-\alpha_0}\label{eq:31}
\end{equation}
and
\begin{equation}
\left[m\right]_{f}! \equiv \left[m\right]_{f} 
\left[m-1\right]_{f}...\left[1\right]_{f} \label{eq:32}.
\end{equation}

The aim herein is to exhibit a JS map that connects
two independent GHA, with the same non-linear characteristic function $f$,
to the G-$sl(2)$ algebra with a non-linear characteristic
function $g$. 
Similarly to the standard JS map, we introduce
the following map
\begin{eqnarray}
S_z &\equiv& G(N_1,N_2) 
\label{eq:33}\\
S_+ &\equiv& F(N_1,N_2) A^\dagger_{1} A_2  \label{eq:34}\\
S_- &\equiv& A^\dagger_{2} A_1 F(N_1,N_2) ,\label{eq:35}  
\end{eqnarray} 
where $N_1$ and $N_2$ are the number operators defined as
\begin{eqnarray}
N_{1} |n_1,n_2\rangle = n_{1} |n_1,n_2 \rangle \label{eq:36} \\
N_{2} |n_1,n_2\rangle = n_{2} |n_1,n_2 \rangle \label{eq:36.1}
\end{eqnarray}
and the functional $F$ and $G$ have the following forms
\begin{eqnarray}
\fl F(N_1,N_2) \equiv \frac{\sqrt{-Q_{2} 
\left[j+1-\frac{N_1-N_2}{2}\right]_{g} \left( 2 \alpha_j +1 + Q_{2} 
\left[ j+1-\frac{N_1-N_2}{2}\right]_{g} \right) }}
{M^{2}_0 \sqrt{  \left[j+1-\frac{N_1-N_2}{2}\right]_{f} 
\left[j+\frac{N_1-N_2}{2} \right]_{f}}}
\label{eq:37}
\end{eqnarray}
and
\begin{eqnarray}
G(N_1,N_2) \equiv \alpha_j + Q_{2}
\left[j- \frac{N_1-N_2}{2} \right]_{g}
\label{eq:38}
\end{eqnarray}
where $Q_2 =g(\alpha_j)-\alpha_j$, $\left[m \right]_{f} = \frac{f^{m}(\alpha_0)-\alpha_0}
{f(\alpha_0)-\alpha_0}$ and $\left[m \right]_{g} 
=\frac{g^{m}(\alpha_j)-\alpha_j}{g(\alpha_j)-\alpha_j}$ $\forall m$
positive integer, where $f$ and $g$ are general analytical functions. 
We can notice that for $f(x)=x+1$ and $g(x)=x-1$ the generalised JS
map recovers the form of the standard JS map \cite{sakurai} since
$G(N_1,N_2) \rightarrow \frac{N_1-N_2}{2}$ and $F(N_1,N_2) \rightarrow 1$. 

As it will become clear later, the operator
\begin{equation}
S^2 \equiv \frac{1}{2} \left[ S_+ S_- + S_- S_+ + S_z (S_z+1) +g(S_z)
(g(S_z) +1) \right] \label{eq:39}
\end{equation}
is the Casimir operator of the algebra generated by $S_{z}$, $S_{+}$
and $S_{-}$. 

If we identify 
\begin{eqnarray}
n_1 \rightarrow j+m \Longrightarrow j=\frac{n_1+n_2}{2} \label{eq:40}\\
n_2 \rightarrow j-m \Longrightarrow m=\frac{n_1-n_2}{2} \label{eq:41}
\end{eqnarray}
we can easily check from eq.(\ref{eq:30}), that the more 
general eigenstate of the operator $S_z$ is written as
\begin{equation}
|j,m\rangle = \frac{(A^\dagger_1)^{j+m}(A^\dagger_2)^{j-m}}{M^{2j}_0
\sqrt{\left[j+m\right]_{f}!
\left[j-m\right]_{f}!}} |0,0\rangle  \label{eq:42}
\end{equation}
where $m=-j,-j+1,...,0,...,j-1,j$.
Without loss of generality we identify $m \rightarrow j-m$ and
denote $|j,j-m\rangle \equiv |\alpha_j , j-m\rangle$ where $m=0,1,2,...,2j$.
The representation of the algebra is fixed by $S_{z}$, $S_{+}$
and $S_{-}$ obtained by the action of these operators on the $S_z$ eigenstates, 
$|\alpha_j,j-m\rangle$, given by: 
\begin{equation}
\fl S_z |\alpha_j,j-m\rangle = (\alpha_j + Q_{2} \left[m
 \right]_{g}) |\alpha_j,j-m\rangle   \label{eq:43}
\end{equation}
\begin{equation} 
\fl S_+ |\alpha_j,j-m\rangle = \sqrt{-Q_{2} 
\left[m\right]_{g} (2 \alpha_j +1 + Q_{2} 
\left[ m\right]_{g})}
 \, \, |\alpha_j,j-m+1\rangle  \label{eq:44} 
\end{equation}
\begin{equation}
\fl S_- |\alpha_j,j-m\rangle =  \sqrt{-Q_{2} 
\left[m+1\right]_{g} (2 \alpha_j +1 + Q_{2} 
\left[ m+1\right]_{g})}
\, \, |\alpha_j,j-m-1\rangle \label{eq:45}
\end{equation}
\begin{equation}
\fl S^2 |\alpha_j,j-m\rangle = \alpha_j (\alpha_j+1) |\alpha_j,j-m\rangle,   \label{eq:46}
\end{equation} 
valid for general analytical functions $f$ and $g$, where we can see
that $S^2$ is the Casimir operator of the algebra. It is very important to
stress that this representation (\ref{eq:43}-\ref{eq:46}) is valid for all
$m=0,1,2,...,2j$. So, the generalised JS map (\ref{eq:33}-\ref{eq:35})
is well defined. 

Once presented the generalised JS map, eqs.(\ref{eq:33}-\ref{eq:35}), we can readily prove that this
map in fact realises the G-$sl(2)$ algebra. 
For this, we will show that the operators $J_0$, $J_+$,$J_-$ and $\widehat{C}$ act on 
the $2j+1$ states of the representation of the G-$sl(2)$
algebra exactly as the operators $S_z$, $S_+$, $S_-$ and $S^2$,
defined in eqs.(\ref{eq:33}-\ref{eq:35},\ref{eq:39}), act 
on the $2j+1$ states of their own representation.
In order to check this, we use $\left[m \right]
_{g} =\frac{g^{m}(\alpha_j)-\alpha_j}{g(\alpha_j)-
\alpha_j}=\frac{\alpha_{j-m}-\alpha_j}{Q_2}$ and $\alpha_{j-m} =
g^{m}(\alpha_j)$. So, we can rewrite the representation of the
G-$sl(2)$ algebra (\ref{eq:18}-\ref{eq:21}) as  
\begin{equation}
\fl J_0 |\alpha_j,j-m\rangle = (\alpha_j + Q_{2} \left[m
 \right]_{g}) |\alpha_j,j-m\rangle   \label{eq:47}
\end{equation}
\begin{equation} 
\fl J_+ |\alpha_j,j-m\rangle = \sqrt{-Q_{2} 
\left[m\right]_{g} (2 \alpha_j +1 + Q_{2} 
\left[ m\right]_{g})}
 \, \, |\alpha_j,j-m+1\rangle  \label{eq:48} 
\end{equation}
\begin{equation}
\fl J_- |\alpha_j,j-m\rangle =  \sqrt{-Q_{2} 
\left[m+1\right]_{g} (2 \alpha_j +1 + Q_{2} 
\left[ m+1\right]_{g})}
\, \, |\alpha_j,j-m-1\rangle \label{eq:49}
\end{equation}
\begin{equation}
\fl \widehat{C} |\alpha_j,j-m\rangle = \alpha_j (\alpha_j
+1)|\alpha_j,j-m\rangle  \label{eq:50}
\end{equation} 
and our proof is complete.

In (\ref{eq:33}-\ref{eq:35}) we see that the generators of the
G-$sl(2)$ algebra are written in terms of the generators of
the GHA. Since there is an interpretation for the GHA in terms of
composite particles, the generalised JS map allows us to carry out this
assertion in such away that the G-$sl(2)$ algebra could be
related to composite particles with angular momentum.  
So, the generalised JS map (\ref{eq:33}-\ref{eq:35}) can be an importante
tool that gets up the physical interpretation for the G-$sl(2)$ algebra, 
which certainly will describe phenomenologicaly several important
features of systems formed by composite particles with angular momentum.

Now, let us discuss the most simple non-linear case where the
characteristic functions $f$ and $g$ are polynomial quadratic ones. In this case,
we would like to stress that it is possible to
simplify the generalised JS map by special choices of the functions 
$f$ and $g$. So, if we assume $f(\alpha_0)= t \alpha_0^{2} 
+ q \alpha_0 +1$, $g(\alpha_j)= -t \alpha_j^{2} + q \alpha_j -1$
and $\alpha_j = - \alpha_0$ we can see that $-Q_2 \left[m \right]_
{g} = M^2_0 \left[m \right]_{f}$
$\forall m$ integer, in such way that the expression for the functional 
$F$ leads to
\begin{eqnarray}
F(N_1,N_2) &=& \frac{\sqrt{
2 \alpha_j +1 + Q_{2} 
\left[ j+1-\frac{N_1-N_2}{2}\right]}}
{\sqrt{ -Q_{2}\left[j+\frac{N_1-N_2}{2}\right]}}
\label{eq:51}
\end{eqnarray}
where, hereafter, $\left[m \right]=\frac{g^{m}(\alpha_j)-
\alpha_j}{g(\alpha_j)-\alpha_j}$ for the above defined quadratic
function $g$. Note also that the representation
(\ref{eq:43}-\ref{eq:46}) is invariant. 

The next step is to check if the above simplified assumptions 
for $f$, $g$ and for the parameters $\alpha_0$
and $\alpha_j$ are possible in the GHA and in the G-$sl(2)$ algebra.

As stated in sections 2 and 3, the representations of each algebra can be studied through the
analyses of the stability of the attractors of $f$ (or $g$) and 
their composed functions.
Similarly discussed in sections 2 and 3, we will consider the
characteristic functions $f(\alpha)= t \alpha^{2} 
+ q \alpha +1$ with $\alpha_0>-q/{2t}$ and $g(\alpha)= -t
\alpha^{2} + q \alpha -1$ with $\alpha_j<q/(2t)$. 
\begin{figure}
\begin{center}
\includegraphics*[angle=0,scale=0.3]{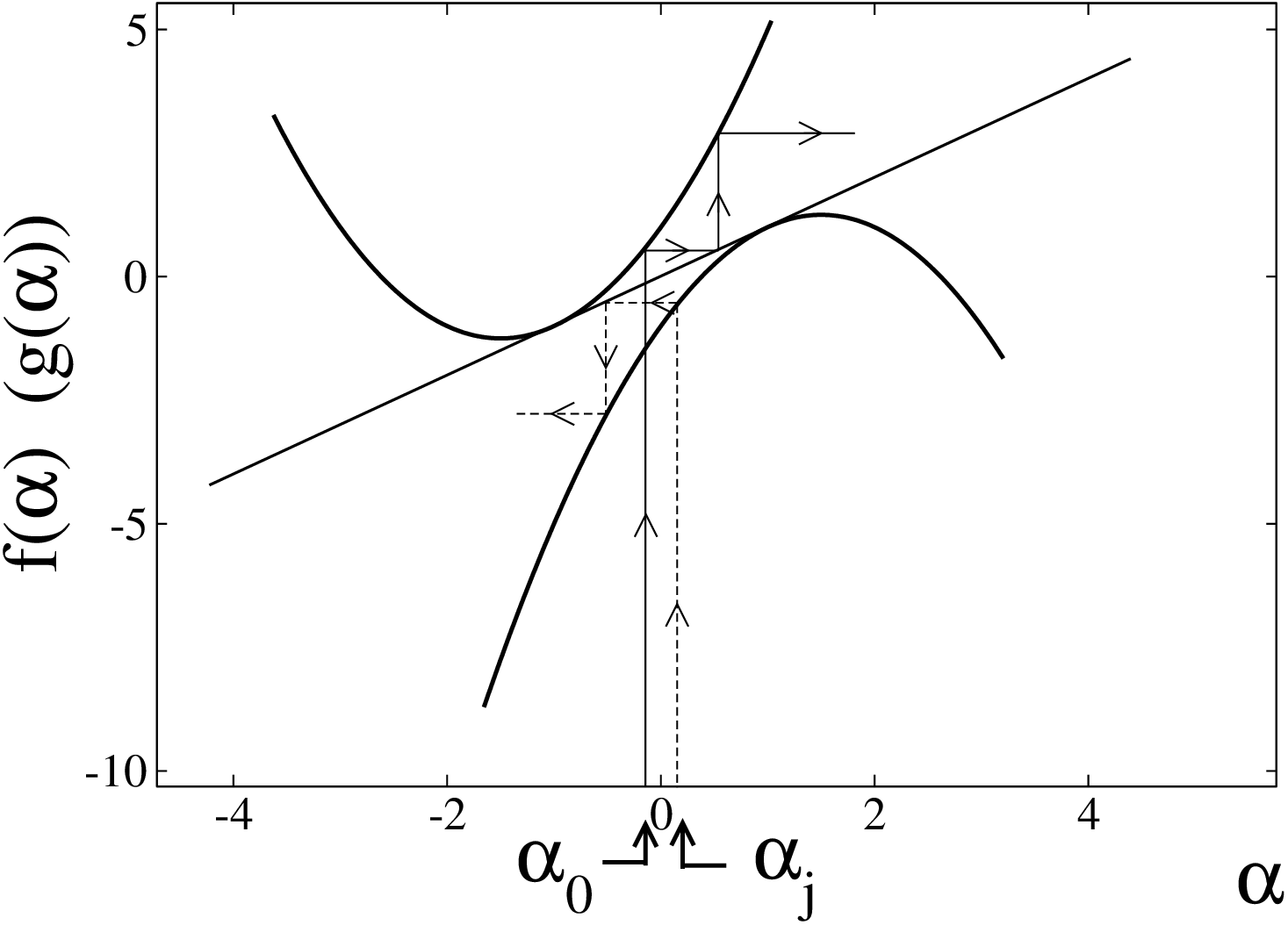}
\caption[]{\small{Graphical analyses of the representations of the G-$sl(2)$ algebra
    and of the GHA for $f(\alpha)=\alpha^2 + 3 \alpha +1$
    and $g(\alpha)=-\alpha^2 + 3 \alpha -1$ ($\Delta=0$). Note that
    there is one fixed point $\alpha^{*}=-1$ and
    $\overline{\alpha^{*}}=1$ for $f$ and $g$, respectively. Note also that it is 
possible to choose $\alpha_0$ as a lowest weight of the GHA
representation that implies $\alpha_j = -\alpha_0$ being the
highest weight of the representation of the G-$sl(2)$ algebra. In this
    figure we show the case $\alpha_0 = -0.15$ and
$\alpha_j =-\alpha_0 =0.15$.

 }}
\label{Fig. 4}
\end{center}
\end{figure}
We see that in the GHA, for $\Delta=(q-1)^2-4t=0$,  
the regions for $\alpha_0$ associated with infinite-dimensional representations 
are given by: $(i)$ $-q/(2t)< \alpha_0 < \alpha^*$ and $(ii)$
$\alpha^* < \alpha_0 < \infty$
where $f^{'}(-q/(2t))=0$ and $\alpha^*$ is the fixed point of $f$. On the other hand, in the G-$sl(2)$
algebra, there are also regions for $\alpha_j$
where we can find a highest weight state and where the algebra
presents infinite-dimensional representations. These regions are:
($a$) $\overline{\alpha^*} < \alpha_{j} < q/(2t)$ 
and ($b$) $-\infty < \alpha_j < \overline{\alpha^*}$,
where $g^{'}(q/(2t))=0$ and $\overline{\alpha^*}$ is the fixed point of $g$.
From the allowed regions for $\alpha_0$ we can see that each value
within it corresponds to a value $\alpha_j =- \alpha_0$ within the
allowed regions for $\alpha_j$ since
\begin{equation}
\overline{\alpha^*} = - \alpha^* \label{eq:52} .
\end{equation}
An example of this case is shown in figure 4 where we plot the functions
$f(\alpha) = \alpha^2 +3 \alpha +1$,
$g(\alpha) = - \alpha^2 +3 \alpha -1$ and $y=\alpha$ versus
$\alpha$. In this figure we exhibit the iterations of $\alpha_0=-0.15$
and $\alpha_j=-\alpha_0=0.15$. The algebras can present also a one-dimensional
representations for $\alpha_0=\alpha^{*}$ and $\alpha_j = -\alpha_0
=\overline{\alpha^{*}}$.

Now we address our atention to discuss the finite-dimensional
representations of the G-$sl(2)$ algebra coming from the cut
condition equation
\begin{equation}
\alpha_j +\alpha_{j-d}+1 =0. \label{eq:56}
\end{equation}

As showed in section 3, for the characteristic function
$g(\alpha)=- \alpha^{2}+3 \alpha -1$, in order to obtain a
two-dimensional representation of the G-$sl(2)$ algebra we
can choose $\alpha_j=0.33479$ or $\alpha_j =2.9228$. In figure 3 we
have showed the iterations for the choice $\alpha_j=0.33479$. We can see that
the value $\alpha_0 = -\alpha_j=-0.33479$ is within the allowed region
for $\alpha_0$, $\alpha^{*}<\alpha_0<\infty$, that corresponds to
an infinite-dimensional representation
for the GHA for $f(\alpha)=\alpha^{2}+3\alpha +1$.

Finaly, we argue that for any general non-linear characteristic functions 
$f$ and $g$, we can always choose, for a given function $f(\alpha_0) = 
\sum_{i=1}^{n} a_i \, (\alpha_0)^{i} +1$, a function $g(\alpha_j) = 
\sum_{i=1}^{n} \overline{a_i} \, (\alpha_j)^{i} -1$ where
$\overline{a_i}=-a_i$, if $i$ is even, and $\alpha_j =- \alpha_0$ in such
away that $-Q_2 \left[m \right]_{g} = M^2_0 
\left[m \right]_{f}$ $\forall m$ integer. So, we can simplify the
functional $F$ in order to turn the generalised JS map more simple. Of course,
given the algebraic parameters we must be careful and to find out the allowed
regions for $\alpha_0$ and $\alpha_j$.

\section{Conclusion}
We introduced a class of generalised JS maps 
which realises the generators of the G-$sl(2)$ algebra with a
non-linear characteristic function $g$, in terms of the generators
of two independent GHA with a non-linear characteristic function $f$.
We shown that it was necessary to generalise the standard JS map,
putting not only a quadratic form of the ladder operators present in each 
independent GHA, but also joining a functional of the number
operators. We have shown also that this generalised JS map has a simple and
compact form which includes the standard JS map as a particular case. 

The need  
of such generalisation come from the fact that these
two classes of generalisations of the Heisenberg and $sl(2)$ algebras,   
contain, in essence, more general algebraic structures than 
their deformed and non-deformed algebras \cite{curado1}.  
In spite of this, we see that is always possible
to simplify the generalised JS map by special choices of the characteristic 
functions that are present in each generalised algebra.
Of course, this statement have to be analysed with more physical accurency as
discussed at the end of the previous section. 

Finaly, we argue that since there are interpretations in terms of composite particles 
for the GHA, the generalised JS map presented in this work 
is a bridge which suggests a physical 
interpreation for the G-$sl(2)$ algebra in terms of
composite particles with angular momentum. Obviously, a investigation of this assertion is highly
desirable in order to check this interpretation.

\ack
The authors thank  CNPq (Brazil) and Pronex/MCT 
for a partial support. The authors also thanks J. A. Redinz for
manuscript revision.

\end{document}